October 1, 1992 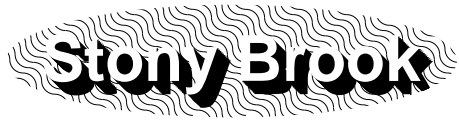 ITP-SB-92-53

# GREEN-SCHWARZ FORMULATION
# OF SELF-DUAL SUPERSTRING

W. Siegel[1]

*Institute for Theoretical Physics*
*State University of New York, Stony Brook, NY 11794-3840*
## ABSTRACT

The self-dual superstring has been described previously in a Neveu-Schwarz-Ramond formulation with local N=2 or 4 world-sheet supersymmetry. We present a Green-Schwarz-type formulation, with manifest spacetime supersymmetry.
---

[1] Work supported by National Science Foundation grant PHY 9211367.
Internet address: siegel@max.physics.sunysb.edu.

# 1. Introduction

The string with local N=2 world-sheet supersymmetry [1] has been shown to describe the self-dual forms of gravity and Yang-Mills theory in two space and two time dimensions (or appropriate dimensional reductions in the heterotic case) [2]. However, the field-theory actions [3] which were originally claimed to correspond to the amplitudes of this string [2] are not Lorentz covariant, even though the field equations of these self-dual theories are Lorentz covariant by definition. Furthermore, these actions defy even dimensional analysis: For example, the Yang-Mills fields used are dimensionless (appearing in exponentials), but they still have the usual d'Alembertian kinetic operator, and thus require a dimensionful coupling constant (in four dimensions), as for nonlinear $\sigma$-models, which they closely resemble [1]. This seems inappropriate for the self-dual restriction of a theory that has a dimensionless coupling because of (spacetime) conformal invariance (classically, or quantum mechanically for the N=4 supersymmetric extension).

The fields used in these actions are related to those appearing in light-cone-like gauges [4,3]. In the past, Lorentz covariance has been an important tool for studying light-cone formulations of strings with or without world-sheet supersymmetry. However, Lorentz transformation properties for these self-dual strings were not considered until recently [5]. There it was found that supersymmetry was an important ingredient in Lorentz covariance: The only field-theory actions that are Lorentz covariant (with the usual vector Yang-Mills field and tensor gravity field, of the right engineering dimensions), give the correct string amplitudes, and have all fields related by supersymmetry ("spectral flow" [6]) are those with maximal supersymmetry.

Besides Lorentz covariance, the major consistency conditions which have always been imposed on string theories involve loop corrections. In the original, nonsupersymmetric descriptions of the self-dual string, there is disagreement between the loop corrections found directly from the string and those from the proposed corresponding field theories [7]. In fact, the string calculations give results characteristic of a two-dimensional field theory (as does the above dimensional analysis of the proposed noncovariant actions, since nonlinear $\sigma$-models have dimensionless couplings only in two dimensions). On the other hand, in the supersymmetric description all loops vanish (at least in the nonheterotic cases) in both string and field theory calculations, and thus the two methods are trivially consistent [5]. (The simplest way to supersymmetrize the usual string calculations is by graded Chan-Paton-like symmetry factors: In light-cone-like gauges, the fields depend on anticommuting coordinates $\theta$, but the action has no explicit $\theta$'s or $\partial/\partial\theta$'s, so the $\theta$-integration in the action causes the component fields, even the fermionic ones, to appear with the same vertices as in the nonsupersymmetric case, but in various combinations.) Finiteness is also believed to be a requirement in general in string theory (the non-finite Veneziano string has problems at least with tachyons and unbounded potentials): The original bosonic version of the self-dual string has loop divergences, while the supersymmetric version is trivially finite because all loops vanish. Since self-dual theories are essentially topological (for example, all tree diagrams except for the three-point function vanish [8,2], even in the bosonic version), loops might be expected to vanish anyway.

Although in [5] we described how to treat the self-dual string theory in a manifestly Lorentz covariant way with the N=4 Neveu-Schwarz-Ramond formalism (the N=2 NSR formalism is just a partial light-cone gauge for the N=4), and how to treat the corresponding field theory in a manifestly Lorentz and supersymmetry covariant way in superspace, we did not describe the string theory in a manifestly spacetime-supersymmetry-covariant way. For the usual superstring, this was achieved [9] by generalizing a formulation of the superparticle [10] in terms of the classical mechanics of the coordinates of superspace. Here we describe the manifestly supersymmetric formulations of the self-dual string, and the equivalent self-dual particle. Because of the unusual nature of self-dual superspace, all constraints are automatically first-class. As a consequence of manifest supersymmetry, external fields can be introduced classically which include all the components of supersymmetric multiplets, allowing nonlinear $\sigma$-model style calculations for all fields, including fermions. Also, the zero-slope "limit", which in this case actually includes all the physical string states, can be taken directly in the classical mechanics string action (unlike NSR formulations of strings), allowing a straightforward transition to the superparticle descriptions of the same theories.

## 2. Particle

Since all the oscillator modes of the self-dual string vanish, its physics can be described by just its zero-modes, the self-dual particle. As in general for relativistic systems, writing a classical mechanics action is equivalent to writing the set of constraints: The lagrangian can be written in hamiltonian form $\dot{q}p - \lambda G(q, p)$ in terms of constraints $G$ with Lagrange multipliers $\lambda$. (For example, for the Veneziano string $G$ is the Virasoro constraints and $\lambda$ the two independent components of the unit-determinant world-sheet metric; the action takes its usual form after eliminating $p$.)

In this paper we'll always work directly in terms of the constraints, but everything can be translated straightforwardly into the language of the classical mechanics action.

Once the nature of the self-dual superspace [5] is understood, the constraints are almost automatic. This superspace has anticommuting spinor coordinates of only one chirality, as might be expected from self-duality: $x^{A\alpha'} = (x^{\alpha\alpha'}, \theta^{a\alpha'})$, where $\alpha, \alpha'$ are the usual spinor indices of SO(2,2)=SL(2,R)⊗SL(2,R) and $a$ is an internal GL(N,R) (Wick rotation of U(N)) index labeling the N spacetime supersymmetries (not to be confused with the world-sheet supersymmetries of the NSR strings). As implied by the indices, this space is a linear realization of a global GL(N|2)⊗GL(2) symmetry. (In self-dual supergravity, an OSp(N|2) subgroup of the GL(N|2) is gauged.) This global symmetry includes not only the SO(2,2) Lorentz and GL(N) internal symmetries, but also scale symmetry, half of the supersymmetry, and half of S-supersymmetry (the "square root" of conformal boosts). In fact, this group is a subgroup of SL(N|4), the superconformal group (which is discussed in more detail below). Besides these manifest "rotational" symmetries of the coordinates $x^{A\alpha'}$, there are also the "translational" symmetries, which include the usual spacetime translations as well as the other half of the supersymmetry. Unlike ordinary superspace, the fermionic translations here are just the naive $\theta \to \theta + \epsilon$; there is no "torsion" in self-dual superspace. Thus, this uncomplicated self-dual superPoincaré group contains not only the usual one but also dilatations and half of S-supersymmetry, and in a simpler form.

The equations of motion (constraints) then follow from the usual Klein-Gordon equation in the unique way which preserves this symmetry:

$$\partial_A{}^{\alpha'} \partial_{B\alpha'} = 0$$

As a consequence of the statistics of the partial derivatives, these equations are graded antisymmetric in the indices $AB$. (We treat the indices $\alpha, \alpha'$ as bosonic, $a$ as fermionic.) Separating these equations into bosonic and fermionic parts,

$$\partial_\alpha{}^{\alpha'} \partial_{\beta\alpha'} = C_{\beta\alpha} \Box = 0$$

$$\partial_\alpha{}^{\alpha'} \partial_{b\alpha'} = 0$$

$$\partial_a{}^{\alpha'} \partial_{b\alpha'} = 0$$

These equations are just the truncation to self-dual superspace of a set of first-class constraints proposed long ago for the superparticle [11]: The first constraint is the Klein-Gordon equation, the second is half of $\slashed{p}d$, the generator of $\kappa$-symmetry [12], and the last is a truncation of *dd*. (In the "chiral representation" half the covariant spinor derivatives can be written as partial derivatives, while the other half, the half which don't explicitly appear in self-dual superspace, take the usual, more complicated form.) In [11] these constraints were shown to follow from superconformal symmetry; here the half of the S-supersmmetry which is manifest is enough to do the job. Just as these covariant derivatives commute (since they're just partial derivatives), these constraints also are abelian. This differs from the usual superspace, where the nonabelian nature of the constraint algebra makes quantization difficult.

These constraints can be solved easily in the light cone. The first constraint is solved as usual; the second just kills half the anticommuting coordinates; the third is then redundant. (For example, in the frame where $\partial_{\alpha-'} = 0$, we find also $\partial_{a-'} = 0$.) In the usual superspace, the first two constraints are not sufficient, since the anticommuting space must be reduced to a quarter the original size to obtain an irreducible representation of supersymmetry; here we already started with half the usual number of fermionic coordinates since the superspace is chiral, so a further division in half completes the process.

## 3. Superconformal group

The open self-dual string describes self-dual N=4 super Yang-Mills theory, which is superconformally invariant. The usual (super)spacetime representation of the superconformal group can be derived by starting with six-dimensional superspace and imposing covariant constraints which reduce the spacetime to four dimensions. In particular, both the self-dual and the non-self-dual representations can be derived from the same six-dimensional superspace and constraints. Essentially, this is because the four-dimensional superconformal group SL(N|4) has a unique linear realization that includes six spacetime coordinates: In terms of SL(N|4) indices $\mathcal{A}$, $\mathcal{B}$,..., the coordinates are graded antisymmetric tensors $x^{\mathcal{AB}}$. This follows from the fact that an antisymmetric tensor of the subgroup SL(4) is a vector of SO(3,3). We also introduce SL(N|4) spin operators $M_\mathcal{A}{}^\mathcal{B}$ ($M_\mathcal{A}{}^\mathcal{A} = 0$). The complete SL(N|4) generators are then

$$J_\mathcal{A}{}^\mathcal{B} = x^{\mathcal{BC}}\partial_{\mathcal{CA}} - \text{trace} + M_\mathcal{A}{}^\mathcal{B}$$

The constraints that eliminate the extra two spacetime dimensions are

$$x^{[\mathcal{AB}}x^{\mathcal{CD})} = 0$$

$$x^{\mathcal{C}[\mathcal{A}}M_\mathcal{C}{}^{\mathcal{B})} + \tfrac{N-2}{N-4}kx^{\mathcal{AB}} = 0$$

$$\tfrac{1}{2}x^{\mathcal{AB}}\partial_{\mathcal{BA}} = 0$$

(The factor "$\frac{N-2}{N-4}$" makes the scale weight N-independent. For N=4, where the superconformal group is actually SSL(N|4), we either set $k=0$ or drop a certain trace part of the generators from the group.) In the case N=0, the first constraint is just the scalar equation $x^2 = 0$. To reduce to four dimensions, we expand $\mathcal{A} = (A, \alpha')$ for the self-dual representation, or further expand $\mathcal{A} = (a, \alpha, \alpha')$ for the usual left-right symmetric representation. For the self-dual case, the constraints become

$$-x\partial + \tfrac{1}{2}x^{AB}\partial_{BA} = 0$$

$$xM^{\alpha'}{}_{\alpha'} + x^{A\alpha'}M_{A\alpha'} + \tfrac{N-2}{N-4}kx = 0$$

$$xx^{AB} - x^{A\alpha'}x^{B}{}_{\alpha'} = 0$$

$$xM^{\alpha'B} + x^{C\alpha'}M_C{}^B + x^{B\gamma'}M_{\gamma'}{}^{\alpha'} + x^{BC}M_C{}^{\alpha'} - \tfrac{N-2}{N-4}kx^{B\alpha'} = 0$$

$$-x^{[A\gamma'}M_{\gamma'}{}^{B)} + x^{C[A}M_C{}^{B)} + \tfrac{N-2}{N-4}kx^{AB} = 0$$

$$x^{[AB}x^{C)\alpha'} = 0$$

$$x^{[AB}x^{CD)} = 0$$

where we have written $x^{\alpha'\beta'} = C^{\beta'\alpha'}x$, $\partial_{\alpha'\beta'} = C_{\beta'\alpha'}\partial$. These constraints, which are also gauge generators, yield the solutions and gauge conditions

$$\partial = 0, \quad x = 1$$

$$M_{\alpha'}{}^{\alpha'} = -M_A{}^A = \tfrac{N-2}{N-4}k$$

$$x^{AB} = x^{A\alpha'}x^B{}_{\alpha'}, \quad \partial_{AB} = 0$$

$$M_{\alpha'}{}^A = -x^B{}_{\alpha'}M_B{}^A - x^{A\beta'}M_{\beta'\alpha'} + \tfrac{N-2}{N-4}kx^A{}_{\alpha'}, \quad M_A{}^{\alpha'} = 0$$

and the last three constraints are redundant. The remaining variables are now just the coordinates $x^{A\alpha'}$ (and their conjugates) of self-dual superspace, the spin operators $M_A{}^B$ and $M_{(\alpha'\beta')}$ of SL(N|2)⊗SL(2), and the scale weight $k$. (As usual for supergroups, for the case N=2 SL(N|2) becomes SSL(N|2), and the traces of both the GL(N) and GL(2) subgroups vanish.) The superconformal generators are now obtained by substituting these results into the original six-dimensional expressions:

$$J_{A\alpha'} = \partial_{A\alpha'}$$

$$J_A{}^B = -x^{B\alpha'}\partial_{A\alpha'} + \widetilde{M}_A{}^B + \tfrac{1}{N-4}k\delta_A{}^B, \quad \tilde{J}_{\alpha'}{}^{\beta'} = -x^{A\beta'}\partial_{A\alpha'} - \text{trace} + \widetilde{M}_{\alpha'}{}^{\beta'}$$

$$J_{\alpha'}{}^{\alpha'} = x^{A\alpha'}\partial_{A\alpha'} + \tfrac{N-2}{N-4}k$$

$$J^{\alpha'A} = x^{A\beta'}x^B{}_{\beta'}\partial_B{}^{\alpha'} - x^{B\alpha'}\widetilde{M}_B{}^A - x^{A\beta'}\widetilde{M}_{\beta'}{}^{\alpha'} + \tfrac{1}{2}kx^{A\alpha'}$$

where "$\widetilde{\phantom{x}}$" means the traceless part. This result is essentially the same as the usual representation of the superconformal group in chiral superspace [13], but simplified by the use of SL(N|2) notation to the point where it looks the same as the representation of the ordinary conformal group (in spinor notation).

## 4. $\sigma$-models

In the $\sigma$-model approach to introducing interactions to strings, external fields are added to the classical mechanics action, and their equations of motion follow upon quantization. For the usual Green-Schwarz superstring, and for the N=2 Neveu-Schwarz-Ramond superstring, some of these equations follow already at the classical level. For the self-dual GS superstring, and for the N=4 NSR string, all field equations appear classically. Here we present the calculations for the last three cases, but introducing the external fields into the constraints instead of the action. This is equivalent, since gauge invariance of the action requires closure of the constraint algebra. (In these cases, unlike the usual GS string, all constraints are first-class: i.e. they all generate symmetries, and their algebra closes.) Also, since all three cases describe the self-dual superstring, we can replace the string with the corresponding particle. (For the NSR cases, we consider just the Ramond sector, since the Neveu-Schwarz sector takes the appearance of an ordinary scalar and is uninteresting.)

We consider external (super) Yang-Mills fields, corresponding to the open string. (In the string constraints, the external field couples to the ends of the string.) This interaction is achieved by the usual minimal coupling $\partial \to \nabla = \partial + A$. For the self-dual superparticle, this gauge covariantization of the equations of motion of the previous section produces the algebra

$$G_{AB} \equiv \tfrac{1}{2}\nabla_{[A}{}^{\alpha'}\nabla_{B)\alpha'} \quad \Rightarrow \quad [G_{AB}, G_{CD}\} = -F_{[A[C}G_{D)B)}$$

where to get the algebra to close we had to impose

$$[\nabla_{A\alpha'}, \nabla_{B\beta'}\} = C_{\beta'\alpha'}F_{AB}$$

which are exactly the equations of self-duality of super Yang-Mills theory, written in self-dual superspace [5].

For the particle version of the N=4 NSR string (taken from the zero-modes of the Ramond sector), the analysis is similar, although in this case the physical description is of a Weyl spinor in an external Yang-Mills field (no supersymmetry). The constraints, compared with their self-dual-superparticle analogs, are

$$\begin{array}{cc} \text{N=4 spinning particle} & \text{superparticle} \\ \Box & \Box \\ \Gamma_a{}^{\alpha'}\nabla_{\alpha\alpha'} & \frac{1}{2}\nabla_{[a}{}^{\alpha'}\nabla_{a]\alpha'} \\ \frac{1}{2}\Gamma_{(a}{}^{\alpha'}\Gamma_{b)\alpha'} & \frac{1}{2}\nabla_{(a}{}^{\alpha'}\nabla_{b)\alpha'} \end{array}$$

The main difference between these two sets of constraints is that for the spinning particle the fermionic variables are $\gamma$-matrices, while for the superparticle they're (gauge covariantized) partial derivatives. (For the corresponding strings, they also have different conformal dimension.) For closure of the algebra

$$\{\Gamma_a{}^{\alpha'}\nabla_{\alpha\alpha'}, \Gamma_b{}^{\beta'}\nabla_{\beta\beta'}\} = C_{ab}C_{\alpha\beta}(\Box + F^{\alpha'\beta'}\Gamma^a{}_{\alpha'}\Gamma_{a\beta'})$$

we need $F_{\alpha\beta} = 0$.

The particle version of the N=2 NSR string is described by a restriction of the N=4 constraints (in terms of the same variables) to the subset $\Box$, $\Gamma_+{}^{\alpha'}\nabla_{-\alpha'}$, $\Gamma_-{}^{\alpha'}\nabla_{+\alpha'}$, $\Gamma_+{}^{\alpha'}\Gamma_{-\alpha'}$. To obtain closure of the algebra

$$(\Gamma_+{}^{\alpha'}\nabla_{-\alpha'})^2 = (\Gamma_-{}^{\alpha'}\nabla_{+\alpha'})^2 = 0$$

we require

$$F_{++} = F_{--} = 0$$

In this case, the complete set of self-duality conditions is not required. However, for the string, worldsheet conformal invariance (i.e. closure of the Virasoro algebra with the external fields included) as usual requires that $F$ satisfy the the usual non-self-dual field equations. Using the Jacobi identities for the covariant derivatives $\nabla_{\alpha\alpha'}$, all these conditions then imply $F_{+-} = 0$, and thus $F_{\alpha\beta} = 0$ [14].

## 5. String

Since the constraints for the self-dual superparticle, which followed uniquely from symmetry considerations, are just the self-dual restriction of those of a certain superparticle with only first-class constraints, and in turn the constraints of that superparticle are the zero-modes of the first-class formulation of the Green-Schwarz superstring, it follows that the self-dual superstring should follow from the self-dual restriction of that superstring. That non-self-dual, first-class formulation of the Green-Schwarz superstring was constructed from an affine Lie algebra $D_{\tilde{\alpha}}(\sigma) = \delta/\delta\Theta^{\tilde{\alpha}} + ...$, $P_{\tilde{a}}(\sigma) = (\delta/\delta X^{\tilde{a}} + X'_{\tilde{a}})/\sqrt{2} + ...$, $\Omega^{\tilde{\alpha}}(\sigma) = \Theta'^{\tilde{\alpha}}$ whose structure constants are $\gamma$-matrices. The zero-mode parts of $D$ and $P$ are the usual covariant derivatives of the superparticle. Truncating to self-dual $\Theta$'s, we find a simpler algebra with no structure constants (just as for the superparticle):

$$P_{a\alpha'} = \frac{\delta}{\delta\Theta^{a\alpha'}}, \quad P_{\alpha\alpha'} = \frac{1}{\sqrt{2}}\left(\frac{\delta}{\delta X^{\alpha\alpha'}} + X'_{\alpha\alpha'}\right), \quad P^a{}_{\alpha'} = \Theta'^a{}_{\alpha'}$$

$$[P_{\mathcal{A}\alpha'}(\sigma), P_{\mathcal{B}\beta'}(\tau)\} = i\delta'(\tau - \sigma)C_{\alpha'\beta'}\eta_{\mathcal{A}\mathcal{B}}$$

$$P_{\mathcal{A}\alpha'} = (P_{\alpha\alpha'}, P_{a\alpha'}, P^a{}_{\alpha'}), \quad \eta_{\mathcal{A}\mathcal{B}} = (C_{\alpha\beta}, \delta^b_a, \delta^a_b)$$

The constraints must then be a simple generalization of the Virasoro constraints:

$$G_{\mathcal{A}\mathcal{B}} \equiv \tfrac{1}{2}P_{[\mathcal{A}}{}^{\alpha'}P_{\mathcal{B})\alpha'} = 0$$

(normal ordered) with the Virasoro constraints themselves given by their trace $L = \tfrac{1}{2}\eta^{\mathcal{B}\mathcal{A}}G_{\mathcal{A}\mathcal{B}}$. Their algebra is

$$[G_{\mathcal{A}\mathcal{B}}(\sigma), G_{\mathcal{C}\mathcal{D}}(\tau)\} = \tfrac{1}{2}i\delta'(\tau - \sigma)\eta_{[\mathcal{A}|[\mathcal{C}}(G_{\mathcal{D})|\mathcal{B})}(\sigma) + (\tau))$$

where we have ignored anomaly terms.

These constraints can be solved quantum mechanically in the same way as for the N=2 (or 4) NSR string [15]. The solution is that none of the oscillators contribute to physical states; the spectrum is given by the massless ground states, the self-dual superparticle. The proof is simplest for the case where the GS string has N=1 spacetime supersymmetry, since in that case it has the same number of fermionic variables as the N=4 NSR string. We first compare the expressions for the constraints

| N=4 NSR string | GS superstring |
|---|---|
| $\tfrac{1}{2}P^{\alpha\alpha'}P_{\alpha\alpha'} + \tfrac{1}{2}i\Gamma^{a\alpha'}\Gamma'_{a\alpha'}$ | $\tfrac{1}{2}P^{\alpha\alpha'}P_{\alpha\alpha'} + \tfrac{1}{2}P^{\tilde{a}\alpha'}P_{\tilde{a}\alpha'}$ |
| $\Gamma_a{}^{\alpha'}P_{\alpha\alpha'}$ | $P_{\tilde{a}}{}^{\alpha'}P_{\alpha\alpha'}$ |
| $\tfrac{1}{2}\Gamma_{(a}{}^{\alpha'}\Gamma_{b)\alpha'}$ | $\tfrac{1}{2}P_{(\tilde{a}}{}^{\alpha'}P_{\tilde{b})\alpha'}$ |

where for the GS case $P_{\tilde{a}\alpha} \equiv (P_{a\alpha}, P^a{}_\alpha)$, and the fermionic commutation relations

| N=4 NSR string | GS superstring |
|---|---|
| $\{\Gamma_{a\alpha'}(\sigma), \Gamma_{b\beta'}(\tau)\} = C_{ab}C_{\alpha'\beta'}\delta(\tau - \sigma)$ | $\{P_{\tilde{a}\alpha'}(\sigma), P_{\tilde{b}\beta'}(\tau)\} = \eta_{\tilde{a}\tilde{b}}C_{\alpha'\beta'}i\delta'(\tau - \sigma)$ |

(Here the NSR $a$ index and the GS $\tilde{a}$ index both take 2 values, with $C_{ab} = \sigma_2$ and $\eta_{\tilde{a}\tilde{b}} = \sigma_1$.) By comparing oscillator expansions for these two formulations, we see that the constraint algebra is identical except for the fermionic zero-modes and the mode-dependent normalization of the terms. (Basically, $\Gamma_{+\alpha'}$ and $\Gamma_{-\alpha'}$ act as $\Theta_{\alpha'}$ and $\delta/\delta\Theta^{\alpha'}$, but $\Theta$ appears as $\Theta'$.) By considering the N=2 NSR subset of the N=4 NSR constraints (as described in section 3) and the corresponding subset of the N=1 GS constraints, to get an irreducible set of constraints, the proof that oscillators do not contribute to physical states is the same for the NSR and GS cases. The only difference is in the fermionic zero-modes which remain in the two formalisms for the particle described by the ground states. (In ref. [14] only the Neveu-Schwarz sector was considered whereas here we need to consider the Ramond sector for the two proofs to be parallel. The treatment is essentially the same except for the treatment of the fermionic zero-modes.)

The Virasoro anomaly calculation for this irreducible subset of the N=1 GS constraints is very simple: The bosonic and fermionic variables have the same conformal weight and come in the same number (four each), as do the bosonic and fermionic constraints (two each). Thus, the anomaly cancels.

The solution of the constraints for the N>1 GS superstring is accomplished by considering an N=1 subset of the constraints. This subset is identical to the N=1 constraints not only in its algebra but also in its representation in terms of four (of the total 4N) of the fermionic variables and all four bosonic variables. Thus, solving these constraints (ignoring the remaining 4(N−1) fermionic variables), and the corresponding gauge conditions, is identical to the N=1 case, leaving dependence on only the zero-modes of those 4+4 variables. The remaining fermionic constraints then easily reduce the remaining fermionic variables to their zero-modes. (Eliminating fermionic excitations is easy even in the N=1 case; only the bosonic variables required any effort.) The result is that the self-dual GS superstring, for all N, reduces to the self-dual superparticle. (The same is not true for the D>4 N=4 NSR string, since the above analogy of constraints to the GS superstring no longer holds: the $\alpha'$ index is extended instead of the $a$ index. Thus neither D nor N of the NSR strings bears any relationship to N of the GS strings.) The anomaly calculation is not as simple as the N=1 case because of the further reducibility of the constraints; we did not find an irreducible analog to the N=2 NSR constraints.

Although we have described the (noncovariant) quantization of the self-dual GS superstring at the free level, we have not considered further restrictions resulting from interactions. Presumably this would determine N to take the maximal value [5].